\documentstyle[preprint,aps,eqsecnum]{revtex}
\begin{document}
\draft
\title
{A NONPERTURBATIVE CALCULATION OF BASIC  \\
 CHIRAL QCD PARAMETERS WITHIN ZERO MODES  \\
 ENHANCEMENT MODEL OF THE QCD VACUUM. II}

\author{V. Gogohia, Gy. Kluge and M. Priszny\'ak}
 
\address{ RMKI, Department of Theoretical Physics,
Central Research Institute for Physics, \\
H-1525,  Budapest 114,  P. O. B.  49,  Hungary}

\maketitle
 
\begin{abstract}
Basic chiral QCD parameters (the pion decay constant, the quark
and gluon
condensates, the dynamically generated quark mass, etc) as well as
the vacuum energy density (up to the sign, by definition, the
bag constant) have
been calculated from first principles within a recently proposed
zero modes enhancement (ZME) model of the true QCD vacuum.
Our unique input data was chosen to be the
pion decay constant in the chiral limit as given by the chiral
perturbation theory at the hadronic level (CHPTh).
In order to analyze our numerical results we set a scale by
two different ways. In both cases
we obtain almost the same numerical results for all chiral QCD
parameters. Phenomenological estimates of these quantites as
well as vacuum energy density are in good agreement
with our numerical results. Complementing them by the numerical
value of the instanton contribution to the vacuum energy density,
we predict new, more realistic values for the vacuum energy
density, the bag constant and the gluon condensate.
\end{abstract}

\pacs{ PACS numbers: 11.30 Rd, 12.38.-t, 12.38 Lg and 13.20 Cz.}

\vfill
\eject

\section{Introduction}

  Let us begin the second part of our paper with the discussion of
one of the most interesting feature of (dynamical chiral symmetry
breaking) DCSB. As was underlined in the first part (hereafter
referred to as I), there
are only five independent quantities by means of which all other
chiral QCD parameters should be calculated. For the sake of
convenience, let us write down them together. They are:
\begin{equation}
F^2_{CA} = {3\over {8 \pi^2}}k_0^2z_0^{-1}
             \int^{z_0}_0 dz \,{ zB^2(z_0,z) \over
             {\{zg^2(z) + B^2(z_0,z)\}}} ,
\end{equation}
\begin{equation}
m_d = k_0\bigl\{z_0 B^2(z_0,0)\bigr\}^{-1/2},
\end{equation}
\begin{equation}
{\langle \overline qq \rangle}_0 = -
{3\over {4\pi^2}}k_0^3z_0^{-3/2}{\int}^{z_0}_0 dz\,{zB(z_0,z)},
\end{equation}
\begin{equation}
\epsilon_q = - {3 \over {8 \pi^2}} k^4_0 z^{-2}_0
\int \limits_0^{z_0} dz\, z\, \{ \ln z\left[ z g^2(z) +
B^2(z_0, z)\right] - 2z g(z) + 2\},
\end{equation}
\begin{equation}
\epsilon_g = - {1 \over \pi^2} k^4_0 z^{-2}_0 \times
\left[ 18 \ln (1 + { z_0 \over 6})
       - {1 \over 2} z^2_0 \ln (1 + {6 \over z_0})
       - {3 \over 2} z_0 \right].
\end{equation}
Recall that $g(z)$ and $B^2(z_0, z)$ are explicitly given by
(1.14) and (1.15) of I. It is instructive also along
with them to write down definition (3.13) of I for
DCSB scale, namely
\begin{equation}
\Lambda_{CSBq} = 2 m_d.
\end{equation}
So these final expressions which should
be used to calculate chiral QCD parameters within our approach
depend only on two independent quantities, namely: mass scale
parameter $k_0$ and the constant
of integration of dynamical quark SD equation of motion $z_0$.
However, it follows from (1.2) that information on the
parameter
$z_0$ should be extracted again from $m_d$ and the initial mass
scale parameter $k_0$
itself, which characterizes the region where confinement, DCSB and
other nonperturbative effects begin to play a dominant role (see
below). So the second indepent parameter
$z_0$ is reduced to the pair of the mass
scale parameters $k_0$ and $m_d$.
Despite the fact that in our consideration the initial mass scale
parameter $\mu$ (characterizing the scale of nonperturbative
effects) has been introduced by "hand", such a transformation of
pair of independent parameters $k_0$ and $z_0$ into the pair of
$k_0$ and $m_d$ is
also a direct manifestation of the phenomenon of the "dimensional
transmutation" [1]. This phenomenon occurs whenever a massless
theory acquires masses dynamically and it is a general feature of
spontaneous symmetry breaking in field theories.
 
  Let us emphasize once more that it  generally characterizes
our approach, in order to calculate numerically all chiral QCD
quantities (considered here and others), that one needs
only two independent (free) parameters, both having
significant and clear physical sense. The above mentioned
dynamically generated quark mass $m_d$, playing a role of the
integration constant of the corresponding equation of motion (the
quark SD equation) instead of $z_0$ because of the above
mentioned "dimensional transmutation" and mass
scale parameter $k_0$,  responsible for a scale at which
important nonperturbative effects begin to play a dominant role.
Our calculation scheme is self-consistent because we calculate
$n=5$ independent physical quantities by means of $m=2$ free
parameters having clear physical sense, so condition of
self-consistensy $n > m$ is satisfied. The general behaviour of
all of our parameters as given by the relations (1.1-1.5)
are shown in Figs. 1-5.
 
 Our approach makes it possible to calculate all chiral QCD
parameters (the ones considered here plus others) at any requested
combination of $m_d$ and $k_0$, but in order to
analyse numerical results it is necessury to set a
scale at which it should be done. We set a scale by two, at
first sight, different ways but leading (see below) to almost the
same numerical results within our calculation scheme.
 
Evidently, to set a scale in each case makes it possible to
determine only one of the two free parameters in our calculations.
In
order to determine the second one we use the chiral value of the pion
decay constant obtained by the chiral perturbation theory at
the hadronic level (CHPTh) in Ref. 2, namely
$F^o_{\pi} = (88.3 \pm 1.1) \ MeV$.
This value is chosen as an input data
in our numerical investigation of chiral QCD.
The pion decay constant is a good experimental number since it
is a directly measurable quantity as opposed, for example, to the
quark condensate. For this reason we may
reliably estimate the deviation of the chiral values
of physical quantities, which can not be directly measured,
from their "experimental", phenomenologically determined values,
when the chiral value of the pion decay constant is approximated
by the experimental value.
 
  In the above mentioned
CHPTh ( or equvivalently the effective field theory)
there is a low energy constant $B$, determined by
${\langle \overline qq \rangle}_0 = - F^2 B$
and measures the vacuum expectation value of the scalar densities
in the chiral  limit. It is just this constant that
determines the meson mass expansion in the general case. Indeed,
in leading order (in powers of quark masses
and $e^2$) from CHPTh, one has [3, 4]
\begin{eqnarray}
 M^2_{\pi^+} & = & (m^0_u + m^0_d) B  \\
 M^2_{K^+} & = & (m^0_u + m^0_s) B  \\
 M^2_{K^0} & = & (m^0_d + m^0_s) B .
\end{eqnarray}
Calculating (independently) the constant $B$, one then
will be able to correctly estimate current quark masses
$m^0_u, m^0_d$ and $m^0_s$ by using the experimental values
of meson masses [5] in (1.7-1.9).

\section{Analysis of the numerical data at a
             scale of DCSB at the quark level}

  Let us begin by recalling that there exists a natural
scale within our approach to DCSB. Indeed, at the
fundamental quark level the chiral symmetry is spontaneously broken
at a scale $\Lambda_{CSBq}$ defined by (1.6). Therefore it makes
sense
to analyse our numerical data at a scale where DCSB at the
fundamental quark level occurs. To this end,
it is necessary only to simply identify mass scale parameter $k_0$
with this scale $\Lambda_{CSBq}$, i.e. to put
\begin{equation}
k_0 \equiv \Lambda_{CSBq} = 2 m_d.
\end{equation}
In other words, we will analyse our numerical results at a
scale responsible for DCSB at the fundamental quark level.
Evidently, this
uniquely determines the constant
of integration of the quark SD equation. Indeed, from
(2.1) and on account of (1.2), then it immediately follows that
this constant is equal to $z_0 = 1.34805$.
  From the pion decay constant
in the chiral limit, chosen as input data, and on account
of this value for $z_0$, from (1.1) it yields the numerical
value for $k_0$. This means that all physical
parameters considered in our paper are uniquely determined.
As it was mentioned above, it will be
instructive to explicitly display our numerical results when the
chiral value of the pion decay constant is approximated by
the experimental value advocated in Refs. 6 and 7, namely
$F^o_{\pi} = 92.42 \ MeV$, as well as by the standard value
$F^o_{\pi} = 93.3 \ MeV$.
Results of our calculations are displayed in Table 1.
 
  Let us make a few concluding remarks. To set
a scale by the way described in this section has the
advantage that it
is based on the exact definition (1.6) for a scale of DCSB at
which
analysis of the numerical data must be done. In
general, it is not
obvious that this scale $\Lambda_{CSBq}$ and scale $\Lambda_c$,
at which
quark confinement occurs, should be of the same
order of magnitude. Moreover, the information about $\Lambda_c$
is hidden within this scheme of calculation.
In order to reveal the raison d'etre for
$\Lambda_c$ and its relation to
$\Lambda_{CSBq}$, let us set a scale in the way described in the
next section.

\section{Analysis of the numerical data at the
            confinement scale}

In our approach there exists only
one scale, denoted as $\mu$ or $k_0$ (separating, in
general,
the nonperturbative phase from the perturbative one), which is
responsible for all the nonperturbative effects in QCD at large
distances. If there is a close relation between quark
confinement and DCSB (and we believe that this is so) then the
scale of DCSB at the fundamental quark level (1.6) and the
confinement scale
$\Lambda_c$ should be, at least, of the same order of magnitude.
In other words, in our approach $\Lambda_c$ should be very
close to $\Lambda_{CSBq}$. This is in agreement with Monte Carlo
simulations on the lattice which show that the deconfinement phase
transition and the chiral symmetry restoring phase transition
occur approximately at the same critical temparature [8],
hereby confirming the close intrinsic link between these
nonperturbative phenomena.
 
   Unfortunately, neither the exact value of $m_d$ nor of
$k_0$ is known. For this reason, let us first
reasonably assume that the dynamically generated
quark masses, in any case,
\begin{equation}
 300   \le   m_d \le 400  \ (MeV),
\end{equation}
should hold but otherwise they remain arbitrary.
We believe that this interval covers all possible realistic values
used for and obtained in various numerical calculations.
The second independent parameter $k_0$ should
be varied in the region of $1 \ GeV$ - the characteristic
scale of low energy QCD.
  Varying independently these pairs
of parameters $m_d$ and $k_0$
numerically, one can calculate all chiral QCD parameters with
the above derived formulae.
 
 From the value of the pion decay constant in the chiral
limit, as well as from the range selected first for $m_d$
(3.1) and on account of
(1.1) and (1.2), it follows that the momentum $k_0$ always
should
satisfy the upper and lower boundary value conditions, namely
$691.32 \le k_0 \le 742.68 \ (MeV)$.
The vacuum energy density contributions of the nonperturbative
gluons
(1.5) changes its sign in the range selected for $m_d$ (3.1)
and in this interval for $k_0$.
Therefore it becomes positive and this should not be so
because of
the normalization condition (we normalize perturbative vacuum to
zero). It is easy to show that this is result of that
that the lower bound
chosen for the dynamical generated quark mass in (4.1) is too
low. Indeed, the vacuum energy density (1.5) vanishes
at the critical point $z^{cr}_0 = 1.45076$ (see Fig. 6). Then from
(1.2) calculated at this point, it follows that
\begin{equation}
k_0 \leq 2.26 m_d.
\end{equation}
Using this ineqaulity in additional, the vacuum energy density
(1.5) will always be negative as it should be and it will
become zero only at critical values determined as
$k^{cr}_0 = 2.26 m_d$.
From the chosen interval for $m_d$
(3.1) and the obtained interval for $k_0$, however, it follows
that the ratio between the corresponding lower bounds $k_0/m_d =
691.32/300 = 2.3044$ does not satisfy the above obtained
inequality (3.2),
while this ratio for the corresponding upper bounds $k_0/m_d =
742.68/400
= 1.8567$ satisfies it. This explicitly shows that the lower
bound for $m_d$ in (3.1) was incorrectly chosen.
The exact lower bound for
$m_d$ can be found from the $k^{cr}_0$ as
$742.68 = 2.26 m_d$,
and (3.1) becomes
\begin{equation}
 328.62 \le m_d \le 400  \ (MeV).
\end{equation}
In the range determined by (3.3) and in the above obtained
interval for $k_0$, the vacuum energy
density (1.5) will be always negative because any combination
(ratio) of $k_0$ and $m_d$ from these intervals will satisfy
inequality (3.2). But this is not the whole story yet. A new lower
bound for $m_d$ leads to a new lower bound for $k_0$ as
well. Indeed, combine now this new lower bound (3.3) with the
chiral value of the pion decay constant one obtains
a new lower bound for $k_0$ as well.
 
  As noted above, $k_0$ is regarded
as a momentum which separates the nonperturbative phase (region)
from the perturbative one. In the region
obtained
for $k_0$ the nonperturbative effects, such as quark confinement
and DCSB, begin to play a dominant role. It is a region
determining a scale at which confinement occurs. From now on let
us call this scale for $k_0$ a confinement scale (in the
chiral limit) and denote it $\Lambda_c$. So the final numerical
values for the confinement scale are as follows
\begin{equation}
707 \le \Lambda_c \le 742.68 \ (MeV).
\end{equation}
In intervals determined by (3.3) and (3.4) the vacuum energy
density $\epsilon_g$ will be always negative (see Fig. 7).
 
 It is worth noting that any value for $\Lambda_c$  from
interval (3.4) is possible but not any combination
of $\Lambda_c$ from interval (3.4) and
$m_d$ from interval (3.3) will automatically satisfy the
value of the pion decay constant.
Therefore it is necessary  to adjust values of $m_d$ from
(3.3) for chosen value of $\Lambda_c$ from interval (3.4) and vice
versa (see Fig. 8).
This means that $m_d$ is in close relationship with $\Lambda_c$.
Moreover, completing the above mentioned procedure, one
finds that $\Lambda_c$ is nearly the double of the generated quark
mass $m_d$, i. e. $\Lambda_c \approx  2 m_d$.
This confirms that $\Lambda_c$ and $\Lambda_{CSBq}$ defined by
(1.6) are nearly the same indeed.
In the previous calculation
scheme the adjusting procedure was automatically fulfilled because
of the exact relation (2.1). Thus there is an intimate
relationship between  $\Lambda_{CSBq}$ and $\Lambda_c$ on
the one hand and the double generated quark mass $m_d$ on the
other hand.
 
 The interval (3.4) for possible values
of $\Lambda_c$
along with the new range for $m_d$ (3.3) will uniquely
determine
numerically the upper and lower bounds for all other chiral QCD
parameters considered here. Like in the previous case, our
numerical results are shown in Table 2 (calculation scheme B),
where the shorthand
${\langle{0}|G^2|{0}\rangle}$ stands for the gluon condensate
${\langle{0}|{\alpha_s \over
\pi}G^a_{\mu\nu}G^a_{\mu\nu}|{0}\rangle}$. Our numerical bounds
for the vacuum energy density $\epsilon$ need additional remarks.
We note that the bounds for $\epsilon$
is not the sum of bounds for $\epsilon_q$  and $\epsilon_g$.
The upper and lower bounds for $\epsilon_q$
are achieved at the upper and lower bounds for $m_d$ ($\Lambda_c$)
while for $\epsilon_g$ they are achieved at the lower and
upper bounds of $m_d$ ($\Lambda_c$).
 
  Let us now prove the relation $\Lambda_c \approx  2 m_d$. We
have already learnt that correct values of $k_0$ belong
to the interval for $\Lambda_c$ (3.4). Then identifying
$k_0$ with $\Lambda_c$ in (3.2), one finally obtains
\begin{equation}
\Delta = \pm (-1 + {\Lambda_c \over \Lambda_{CSBq}}) \leq 0.13,
\end{equation}
where the positive sign corresponds to $\Lambda_c >
\Lambda_{CSBq}$
and the negative one is valid when $\Lambda_c < \Lambda_{CSBq}$.
In the derivation of this relation we used definition (1.6).
 
  Finally it is worth underlining once more that besides
good numerical results obtained
in this section, we have established the
existence of realistic lower bound for the dynamically generated
quark masses. In each calculated case their numerical values are
shown in
Table 2. Thus one concludes that the vacuum energy density due to
the nonperturbative gluons is sensitive to the lower bound for
$m_d$. Another important result is that we have
clearly shown that the confinement scale $\Lambda_c$ and DCSB
scale $\Lambda_{CSBq}$ are nearly the same indeed.

\section{Conclusions and Discussion}

Let us briefly compare our numerical results obtained from
first principles with phenomenologically estimated values of
the physical parameters considered here.
An estimate of the quark condensate in Refs. 9 and 10,
\begin{equation}
{\langle \overline qq \rangle}_0^{1/3}  = -(225 \pm  25) \ MeV
\end{equation}
is in good agreement with our values. It is worth noting
here that QCD sum rules give usually the numerical values of
physical quantities, in particular the quark condensate,
approximately
within an accuracy of (10-20)\% (see, for example Ref. 11).
 
   Our values for the current quark masses are also in good
agreement with recent estimates from hadron mass splittings [12]
\begin{eqnarray}
m^0_u = (5.1 \pm 0.9)  \ MeV,   \nonumber\\
m^0_d = (9.0 \pm 1.6)  \ MeV,   \nonumber\\
m^0_s = (161 \pm 28) \ MeV
\end{eqnarray}
and QCD sum rules [13]
\begin{eqnarray}
m^0_u = (5.6 \pm 1.1)  \ MeV,   \nonumber\\
m^0_d = (9.9 \pm 1.1)  \ MeV,   \nonumber\\
m^0_s = (199 \pm 33) \ MeV,
\end{eqnarray}
see also reviews [14].
 
Here it is worth noting
that from our numerical results (Tables 1 and 2)
it follows that the constituent quark mass $m_q$ should differ
little from $m_d$. Apparently, the difference between them is of
the order of a few per cent only of the displayed values of
$m_d$.
So without making a big mistake even for light quarks, it is
possible to simply use $m_d$ instead of $m_q$. Doing so
one comes to the conclusion that the CHPTh value of the pion decay
constant and the constituent quark model (CQM) with the
value for the constituent
quark mass $m_q=362 \ MeV$ advocated by Quigg [15] are nearly in
one-to-one correspondence within our calculation scheme (see
Table 1). Moreover, from our numerical results (Tables 1 and
2) one can conclude that the dominant contributions to
the values of all chiral QCD parameters as well as the vacuum
energy density come from large distances, while the contributions
from the short and intermediate distances can only be treated as
small perturbative corrections.
 
  The phenomenological analysis of the QCD sum rules [10] for the
numerical value of the gluon condensate implies
\begin{equation}
\langle{0}|{\alpha_s \over
\pi}G^a_{\mu\nu}G^a_{\mu\nu}|{0}\rangle \simeq  0.012 \ GeV^4,
\end{equation}
and using then (2.5) of I, one obtains the vacuum
energy density as
\begin{equation}
\epsilon \simeq -0.003375 \ GeV^4.
\end{equation}
In the random instanton
liquid model (RILM) [16] of the QCD vacuum, for a dilute
ensemble, one has
\begin{equation}
\epsilon =-{ 9 \over 4} \times 1.0 \ fm^{-4} \simeq -0.003411 \ GeV^4.
\end{equation}
The estimate of the gluon condensate within the
QCD sum rules approach can be changed within a factor of two [10].
We trust our numerical results for the vacuum energy density
much more than those of the gluon condensate.
The former was obtained on the basis of the completely
nonperturbative ZME model of the vacuum of QCD while the latter
was obtained on account of the perturbative solution for the
CS-GML $\beta$-function [10]. Let us
also emphasize the one important
fact that our calculation of the vacuum energy density is a
calculation from first principles while in the RILM [16] the
parameters
characterizing vacuum, the instanton size $\rho_0 = 1/3 \ fm$
and the "average separation" $R= 1.0 \ fm$ were chosen
to precisely reproduce traditional (phenomenologically estimated
from the QCD sum rules) values of quark and gluon condensates,
respectively.
 
We reproduce values (4.4-4.6), which are due to
the instanton-type fluctuations only, especially well
when the pion decay constant in the chiral limit was
approximated by its experimental value. Moreover, our numerical
results clearly show that the contribution to the vacuum energy
density of the confining quarks with dynamically generated
masses $\epsilon_q$ is approximately equal to $\epsilon_g$, that
is of the nonperturbative gluons. It is
well known that in the chiral limit (massless
quarks) tunneling is totally suppressed, i.e. the contribution
of the instanton-type fluctuations
to the vacuum energy density vanishes. It
will be restored again in the presence of DCSB [17-19]. Thus, in
principle, in the chiral limit and in the presence of DCSB, the
total vacuum energy density should be the sum (as minimum) of
these three quantities, i.e.
\begin{equation}
\epsilon_t = \epsilon_I + \epsilon_g  + N_f \epsilon_q,
\end{equation}
where $\epsilon_I$ describes the contribution of the
instanton component to the vacuum energy density. We introduce
also the explicit dependence on the number of different quark
flavors $N_f$ since $\epsilon_q$ itself is the contribution of a
single confining quark. Of
course, this should be valid for the non-chiral case as well. The
distinction will be in the concrete values of each component,
apart from, maybe, $\epsilon_g$.
 
Let us now run a risk and make a few quantitative predictions.
Indeed, not going into the details of the instanton physics
(well described in Ref. 18) and in agreement with the authors of
Ref. 10, it is worth assuming, for simplicity's sake, that the
light and heavy quarks match smoothly. This allows one to
choose for the
instanton component of the vacuum energy density $\epsilon_I$
the average value between (4.5) and (4.6), i.e. namely
$\epsilon_I \simeq -0.0034 \ GeV^4$. Then our predictions for
more realistic values of the total vacuum energy density
$\epsilon_t$ (for $N_f$ light confining quarks with dynamically
generated masses) and the corresponding values
of the gluon condensate are listed in Tables 3 and 4, 5 for both
calculation schemes A and B, respectively.
 
  It is worth reproducing explicitly some interesting particular
values of the total vacuum energy density and the corresponding
values of the gluon condensate. Thus for a pure gluodynamics
($N_f=0$) one has
\begin{eqnarray}
\epsilon_t &\simeq& - 0.005 \ GeV^4, \nonumber\\
\langle{0}|{\alpha_s \over
\pi}G^a_{\mu\nu}G^a_{\mu\nu}|{0}\rangle &\simeq&  0.0177 \ GeV^4,
\end{eqnarray}
and
\begin{eqnarray}
- 0.00661 &\leq \epsilon_t \leq& - 0.003837 \ (GeV^4), \nonumber\\
0.0136 \leq &\langle{0}|{\alpha_s \over
\pi}G^a_{\mu\nu}G^a_{\mu\nu}|{0}\rangle& \leq  0.0235 \ (GeV^4).
\end{eqnarray}
Here and below the numbers correspond to the approximation of the
pion decay constant in the chiral limit by its standard value.
 
For the more realistic case $N_f=2$ one obtains
\begin{eqnarray}
\epsilon_t &\simeq& - 0.008 \ GeV^4, \nonumber\\
\langle{0}|{\alpha_s \over
\pi}G^a_{\mu\nu}G^a_{\mu\nu}|{0}\rangle &\simeq&  0.0283 \ GeV^4,
\end{eqnarray}
and
\begin{eqnarray}
- 0.00933 &\leq \epsilon_t \leq& - 0.00724 \ (GeV^4), \nonumber\\
0.0256 \leq &\langle{0}|{\alpha_s \over
\pi}G^a_{\mu\nu}G^a_{\mu\nu}|{0}\rangle& \leq  0.0331 \ (GeV^4).
\end{eqnarray}
 There exist already phenomenological estimates of the gluon
condensate [20] as well as lattice calculations of the vacuum
energy density [21] pointing out that the above mentioned standard
values (4.4) and (4.5-4.6) are too small. Our numerical
predictions are in agreement with these estimates though we think
that their numbers for gluon condensate [20] are too big. At the
same time, it becomes quite clear why the standard values are so
relatively small, because they are due to the instanton component
of the vacuum only.
 
In the above mentioned RILM [16], light quarks
can propagate over large distances in the QCD vacuum by simply
jumping from one instanton to the next. Within our model (see
I and Ref. 20) propagation of all quarks is determined by
the corresponding SD equations (due to the ZME effect) so that
they always remain off mass-shell.
Thus we need no picture of jumping quarks. Contrast to
the RILM, we think that the main role of the instanton-like
fluctuations is precisely to prevent quarks and gluons from the
freely propagation in the vacuum of QCD. Running against
instanton-like fluctuations quarks undergo difficulties in their
propagation in the QCD vacuum which, in principle, is a very
complicated inhomogenious medium.
 At some critical value of the instantons density the free
propagation of the virtual quarks, apparently,
become impossible, so they already never annihilate again with
each other. Obviously, this is equivalent to the creation of the
quark-antiquark pairs from the vacuum. From this
moment nontrivial rearrangement of the vacuum can start.
The above mentioned critical value can be reached when
$\epsilon_I \simeq \epsilon_g + \epsilon_q$, i.e. when at least
one sort of quark flavors is presented in the QCD true vacuum.
On one hand, this is supported by our numerical results for
$\epsilon = \epsilon_g + \epsilon_q$. On the other hand, the
numerical values for $\epsilon_I$, as given by (4.5) or (4.6),
also confirms this. In the realistic (nonchiral) case the
instanton part, along with other contributions, may substantilly
differ from those shown in Tables 1, 2, 3, 4 and 5.
 
 The bag constant is defined as the difference between the energy
density of the perturbative and the nonperturbative QCD vacuums.
We normalize the perturbative vacuum to zero (I). So in
our notations the bag constant becomes
\begin{equation}
B = - \epsilon_t
\end{equation}
(Not to be mixed with the CHPTh constant (1.7-1.9)). Our
predictions for this quantity are also shown in Tables 3 and 4, 5
for each calculation scheme A and B, respectively.
In fact, our values for the bag constant overestimate the initial
MIT bag [23] volume energy by one order of magnitude.
Nevertheless, we think that the introduction of this constant
into physics was a main achievement of the bag model.
 
 As in previous case, let us explicitly reproduce some interesting
concrete values of the bag constant. For a pure gluodynamics
($N_f=0$) it is:
\begin{equation}
B \simeq 0.005 \ GeV^4 \simeq (266 \ MeV ))^4 \simeq 0.651 \
GeV/fm^3.
\end{equation}
For the more realistic case $N_f=2$ the bag constant becomes
\begin{equation}
B \simeq 0.008 \ GeV^4 \simeq (300 \ MeV)^4 \simeq 1 \ GeV/fm^3.
\end{equation}
For simplicity's sake we reproduced its value obtained within the
calculation scheme A only. It has been noticed in [24] that
noybody knows yet how big the bag constant might be, but
generally it is thought it is about $1 \ GeV/fm^3$. The predicted
value for $N_f=2$ is in fair agreement with expectation.

\acknowledgments

  The authors would like to thank J. Zim\'anyi, K. Szeg\H o,
Gy. P\'ocsik, K. Lad\'anyi, B. Luk\'acs, K. T¢th, T. Dolinszky,
G. P\'alla, J. R\'evai and other members of the Theoretical
Department
of the RMKI for their constant interest, valuable discussions
and support. One of the authors (V. G.) is also grateful to
V.A. Rubakov, N. Brambilla, W. Lucha,
F. Schoberl for useful discussions and remarks and especially to
N.B. Krasnikov for detailed discussions on the QCD true vacuum
properties and his persistent advice to take seriously the
relation (2.5) of the first part despite its perturbative origin.

\vfill
\eject

\vfill
\eject

\begin{table}
\caption{Calculation scheme A}
 
\begin{tabular}{|l|l|c|r|r|} \hline
$F^0_{\pi}$ & 88.3 & 92.42 & 93.3 & MeV \\ \hline
 
$\Lambda_{CSBq}$ & 724.274 & 758.067 & 765.284 & MeV \\
 
$m_d$ & 362.137 & 379.0335 & 382.642 & MeV \\
 
${\langle \overline qq \rangle}_0$ & $(-208.56)^3$ & $(-218.29)^3$
& $(-220.36)^3$ & $MeV^3$ \\
 
$\epsilon_q$ & $-0.0012$ & $-0.00143$ & $-0.0015$ & $GeV^4$ \\
 
$\epsilon_g$ & $-0.0013$ & $-0.00157$ & $-0.0016$ & $GeV^4$  \\
 
$\epsilon$ & $-0.0025$ & $-0.0030$ & $-0.0031$ & $GeV^4$  \\
 
$\langle{0}|{\alpha_s \over
\pi}G^a_{\mu\nu}G^a_{\mu\nu}|{0}\rangle$ & 0.009 & 0.0106 &
0.011 & $GeV^4$  \\
 
$m^0_u$ & 6.65 & 6.36 & 6.30 & MeV \\
 
$m^0_d$ & 10.08 & 9.63 & 9.54 & MeV \\
 
$m^0_s$ & 202.85 & 193.75 & 191.94 & MeV \\ \hline
\end{tabular}
\end{table}

\vfill
\eject

\begin{table}
\caption{Calculation scheme B}
\begin{tabular}{|l|c|r|} \hline
$F^o_{\pi}$ = 88.3 & $F^o_{\pi}$ =  92.42 & $F^o_{\pi}$ = 93.3
\\ \hline
 
$707 \le \Lambda_c \le 742.68$ & $737.9 \le \Lambda_c \le
768.4$  & $744.4 \le \Lambda_c \le 773.86$  \\
 
$328.62 \le m_d \le 400$ & $340 \le m_d \le 400$ & $342.416 \le
m_d \le 400$ \\
 
$(-210.34)^3 \le {\langle \overline qq \rangle}_0 \le (-206.9)^3$
&$(-219.3)^3 \le {\langle \overline qq \rangle}_0 \le (-216.34)^3$
&$(-221.2)^3 \le {\langle \overline qq \rangle}_0 \le (-218.33)^3$ \\
 
$-0.00135 \le \epsilon_q \le -0.00096$ & $-0.0016 \le \epsilon_q
\le -0.00128$ & $-0.0017 \le \epsilon_q \le -0.00136$ \\
 
$-0.0024 \le \epsilon_g \le -0.00045$ & $-0.00226 \le \epsilon_g
\le -0.00044$ & $-0.00221 \le \epsilon_g \le -0.000437$  \\
 
$-0.00336 \le \epsilon \le -0.0018$ & $-0.00354 \le \epsilon \le
-0.002$ & $-0.00356 \le \epsilon \le -0.0021$ \\
 
$0.0064 \le {\langle{0}|G^2|{0}\rangle} \le 0.0128$
& $0.007 \le {\langle{0}|G^2|{0}\rangle} \le 0.0192$
& $0.00746 \le {\langle{0}|G^2|{0}\rangle} \le 0.0199$ \\
 
$6.48 \le m^0_u \le 6.81$ & $6.27 \le m^0_u \le 6.53$  &
$6.22 \le m^0_u \le 6.47$ \\
 
$9.83 \le m^0_d \le 10.33$ & $9.5 \le m^0_d \le 9.89$ &
$9.43 \le m^0_d \le 9.81$ \\
 
$197.67 \le m^0_s \le 207.7$ & $191 \le m^0_s \le 199$ &
$189.76 \le m^0_s \le 197.34$ \\ \hline
\end{tabular}
\end{table}

\vfill
\eject

\begin{table}
\caption{Calculation scheme A. Predictions}
\begin{tabular}{|l|l|r|r|} \hline
$F^o_{\pi}$ & 92.42 & 93.3 & MeV \\ \hline
 
$\epsilon_t = \epsilon_I + \epsilon_g + N_f \epsilon_q$ &
$-0.00497 - N_f 0.00143$ & $-0.005 - N_f 0.0015$ & $GeV^4$ \\
 
$\langle{0}|{\alpha_s \over \pi}
G^a_{\mu\nu}G^a_{\mu\nu}|{0}\rangle$ & $0.01767 + N_f 0.00508$ &
$0.01777 + N_f 0.00533$ & $GeV^4$  \\
 
B  & $0.00497 + N_f 0.00143$ & $0.005 + N_f 0.0015$ & $GeV^4$  \\
\hline
\end{tabular}
\end{table}

\vfill
\eject

\begin{table}
\caption{Calculation scheme B. Predictions}
\begin{displaymath}
\begin{array}{|c|} \hline
\; F^o_{\pi} = 92.42 \; \\ \hline
\; -0.00566 - N_f 0.00128 \le \epsilon_t
\le -0.00384 - N_f 0.0016 \;
\\
\; 0.0136 + N_f 0.00568 \le {\langle{0}|G^2|{0}\rangle} \le 0.020
+ N_f 0.0045 \;
\\
\; 0.00384 + N_f 0.0016 \le B \le 0.00566 + N_f 0.00128 \;
\\  \hline
\end{array} \nonumber
\end{displaymath}
\end{table}

\vfill
\eject

\begin{table}
\caption{Calculation scheme B. Predictions}
\begin{displaymath}
\begin{array}{|c|} \hline
\; F^o_{\pi} = 93.3 \; \\ \hline
\; -0.00661 - N_f 0.00136 \le \epsilon_t
\le -0.003837 - N_f 0.0017 \;
\\
\; 0.0136 + N_f 0.006 \le {\langle{0}|G^2|{0}\rangle} \le 0.0235
+ N_f 0.0048 \;
\\
\; 0.003837 + N_f 0.0017 \le B \le 0.00661 + N_f 0.00136 \;
\\  \hline
\end{array} \nonumber
\end{displaymath}
\end{table}

 \vfill
 \eject

\begin{figure}

\caption{The pion decay constant $F_{CA}$
as a function of $k_0$, drawn only for the most reasonable region,
selected first for the dynamically generated quark masses (3.1).
The obtained interval for $k_0$ is also explicitly shown (see
Section 3 below).}

\bigskip
 
\caption{The quark condensate
as a function of $k_0$, drawn only for the most resonable region,
selected first for the dynamically generated quark masses (3.1).}
 
\bigskip
 
\caption{The vacuun energy density due to confining quarks with
dynamically generated masses,
as a function of $k_0$, drawn only for the most resonable region,
selected first for the dynamically generated quark masses (3.1).}
 
\bigskip
 
\caption{The vacuum energy density due to the
nonperturbative gluons
as a function of $k_0$, drawn only for the most resonable region,
selected first for the dynamically generated quark masses (3.1).
The obtained interval for $k_0$ is also explicitly shown (see
Section 3 below).}
 
\bigskip
 
\caption{The vacuun energy density $\epsilon$
as a function of $k_0$, drawn only for the most resonable region,
selected first for the dynamically generated quark masses (3.1).}
 
\bigskip
 
\caption{The vacuum energy density due to the
nonperturbative gluons contributions (1.5) as a function
of $z_0$.}
 
\bigskip
 
\caption{The vacuum energy density due to the nonperturbative
gluon
contributions $\epsilon_g$ as a function of $k_0$. $\Lambda_c$ is
the confinement scale (3.4). A new interval for $m_d$ (3.3) is
also
shown. A similar figure can be drawn for the case when the pion
decay constant is approximated by the experimental value. }

\bigskip
 
\caption{The pion decay constant $F_{CA}$
as a function of $k_0$. $\Lambda_c$ is
the confinement scale (3.4). A new interval for $m_d$ (3.3) is
also
shown. A similar figure can be drawn for the case when the pion
decay constant is approximated by the experimental value. }
\end{figure}

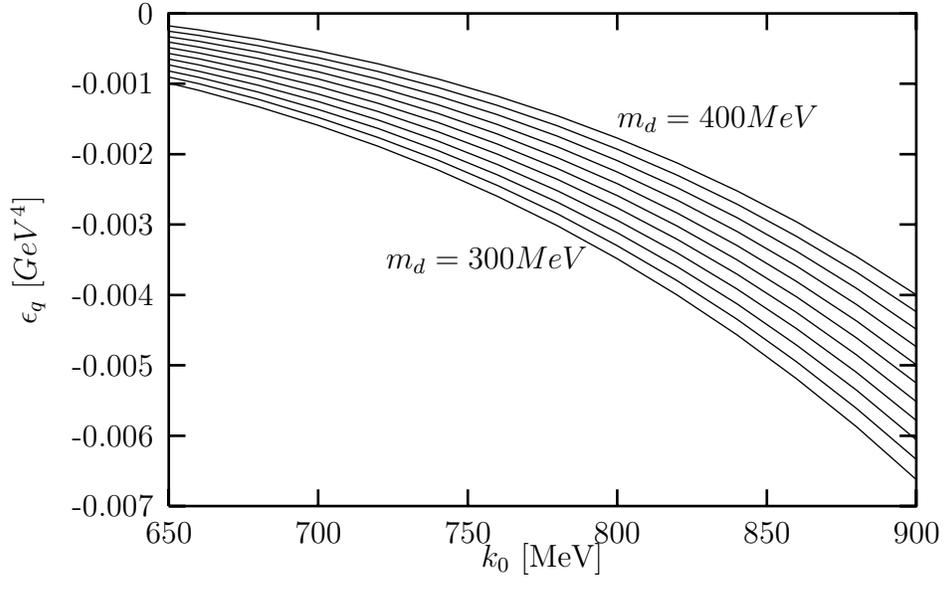
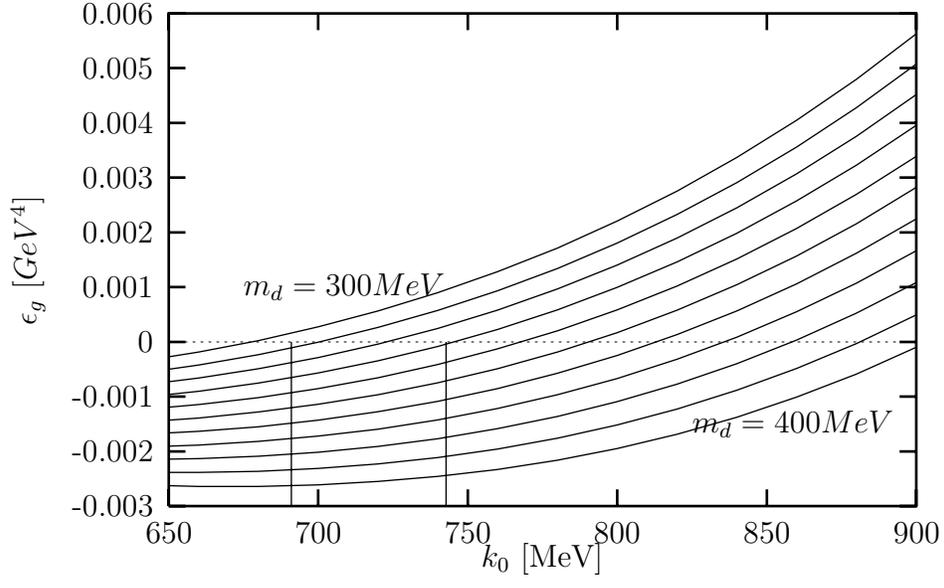
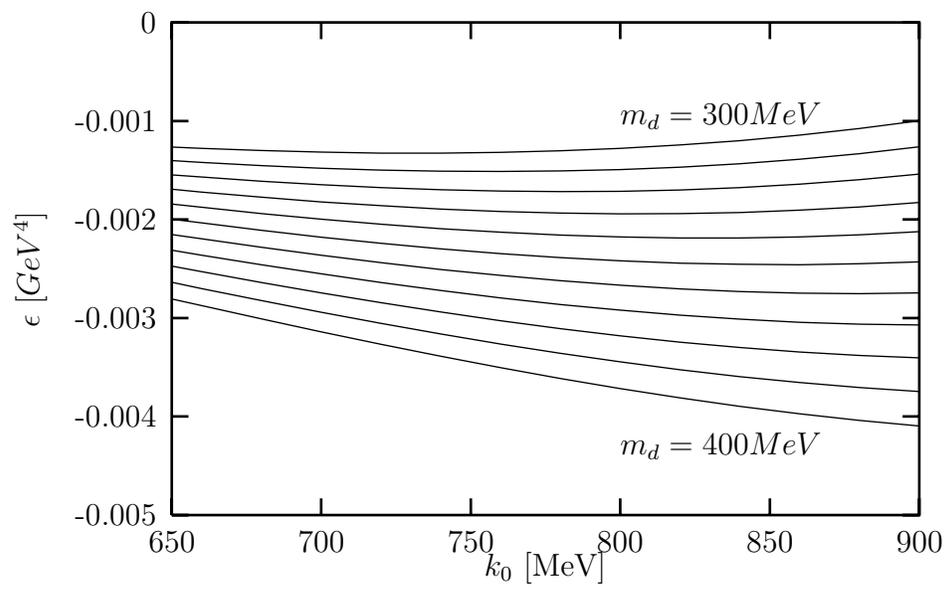
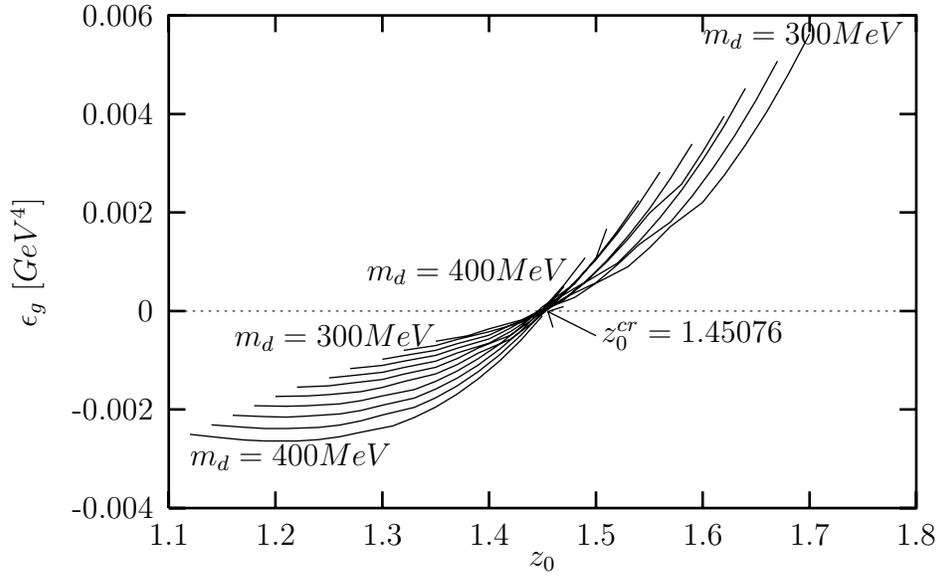
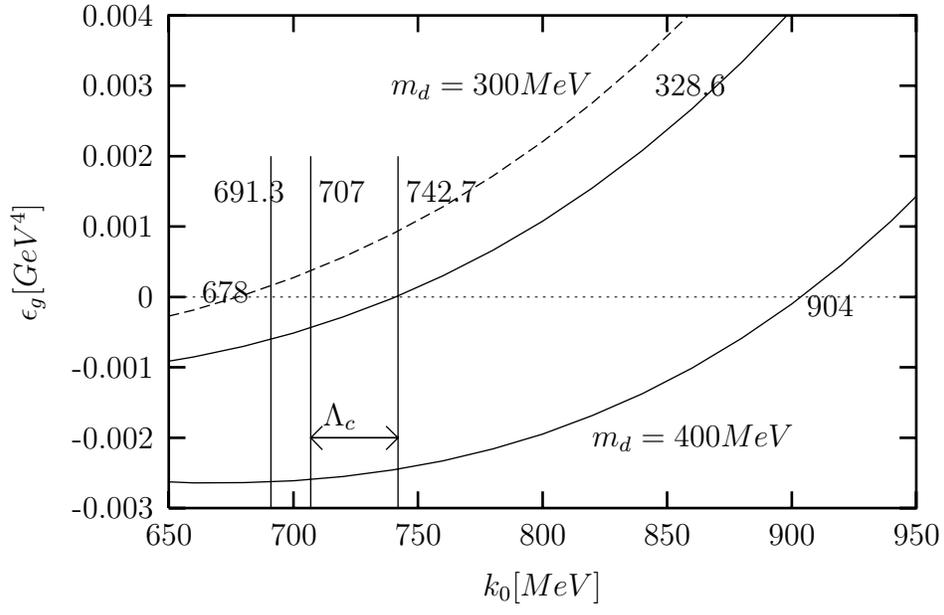
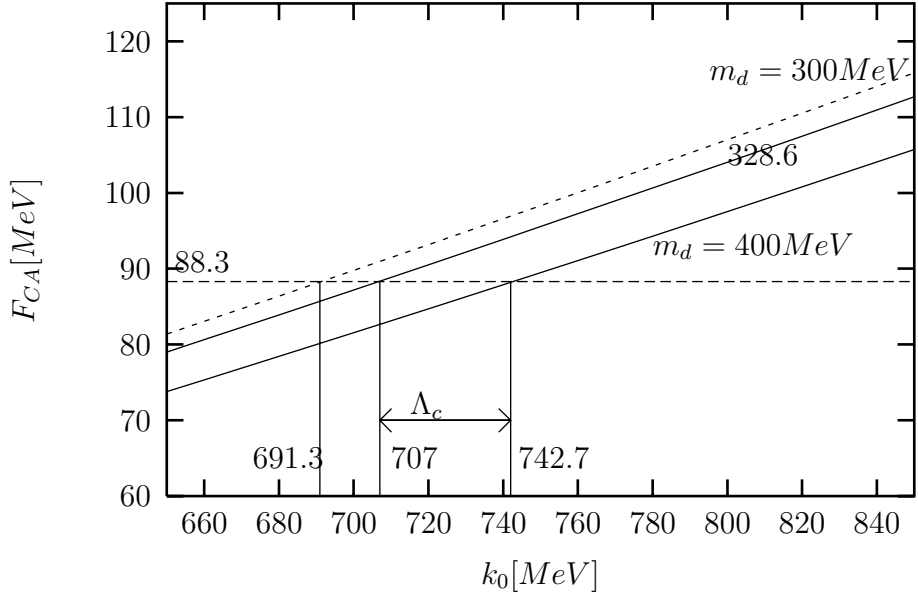
\end{document}